\documentstyle[12pt] {article}
\begin{document}
\baselineskip 28 pt plus2pt
\title {Are physical objects necessarily burnt up by the blue sheet
inside a black hole?}
\author {Lior M. Burko and Amos Ori\\
Department of Physics\\
Technion - Israel Institute of Technology\\
 32000 Haifa, Israel}
\maketitle
\begin{abstract}
The electromagnetic radiation that falls into a Reissner-Nordstr\"{o}m
black hole develops
a ``blue sheet'' of infinite energy density at the Cauchy horizon.
We consider
classical electromagnetic fields (that were produced during the collapse
and then backscattered into the black hole), and
investigate the blue-sheet effects of these fields on infalling
objects within a simplified model.
These effects are found to be finite and even negligible for
typical parameters.
%
%\pacs{04.70.Bw  04.40.Nr}
\end{abstract}
The Reissner-Nordstr\"{o}m spacetime is the unique electrically charged,
spherically
symmetric, static vacuum solution of the Einstein-Maxwell equations.
Although astrophysical black holes are not likely to be significantly
charged, the Reissner-Nordstr\"{o}m solution (which is simple to deal
with due to its
spherical symmetry) can serve as a toy model for more realistic black
holes, such as Kerr. This may be
physically justified by the similarity of the inner causal structures
of the Reissner-Nordstr\"{o}m and Kerr solutions -- as expressed, e.g.,
in similar conformal Penrose diagrams \cite{hawking}.

Even though the causal structure of a Reissner-Nordstr\"{o}m
black hole admits a hypothetical
journey into another asymptotically flat Universe \cite{novikov}, it
turned out that undertaking such a journey might actually be dangerous
\cite{penrose}.
The ``tunnel'' inside the black hole is crossed by a null hypersurface
known as the inner horizon. This null hypersurface is also a Cauchy
horizon, i.e., it is the boundary of the domain of dependence for initial
data specified on spacelike hypersurfaces in the external Universe.
Electromagnetic (or gravitational) radiation
falling into the black hole becomes infinitely blue-shifted at the Cauchy
horizon, typically causing the energy density to blow up
\cite{penrose,gursel,chandrasekhar1}.
This could cause two kinds of problems: First, the divergent flux of
energy carried by the electromagnetic waves might heat any infalling
physical object unboundedly, thus burning it up. Second, acting as a
source term in the Einstein equations, the divergent energy density leads
to a divergent curvature at the Cauchy horizon.
The infalling object would then
experience an unlimited tidal force, which might lead to its ultimate
destruction due to the tidal distortion. (A more direct cause for the
diverging curvature at the Cauchy horizon is the infinite blue-shift
of the gravitational waves, which leads to the divergence of the gradient
of the metric perturbations.)

Recently, there has been growing evidence \cite{ori}, that the
divergence of the curvature at the Cauchy horizon is rather weak. Namely,
the actual tidal deformation suffered by a physical object as it reaches
the Cauchy horizon is finite, and (for typical parameters) even
negligible. If this is indeed the case, the ability of physical objects
to traverse the Cauchy horizon may depend crucially on the other
potential problem, i.e., the possible annihilation due to the divergent
electromagnetic radiation. The main goal of this letter is to investigate
this issue within a simplified model.

In what follows, we consider an isolated charged black hole,
surrounded by electromagnetic waves, which we treat as a linear
perturbation. (In fact, because of the non-vanishing electric field of
the background, this linear perturbation consists of both electromagnetic
and gravitational waves \cite{gursel,chandrasekhar1}.)
First, we calculate the asymptotic behavior of
the electromagnetic perturbation near the Cauchy horizon. Then, we use a
simplified model to evaluate the possible effects
of this field on (test-) infalling objects. Let us denote by $M,Q_{*}$
the mass and charge of the black hole, and by $r$ the radial Schwarzschild
co-ordinate. Let $\Delta =r^{2}-2Mr+Q_{*}^{2}$.
The horizons of the Reissner-Nordstr\"{o}m black hole are the event horizon
$r_{+}$ and the inner horizon $r_{-}$, which are located at the roots of
$\Delta$, namely, at $r_{\pm}=M\pm (M^{2}-Q_{*}^{2})^{1/2}$.
We define the null co-ordinates $u=r_{*}-t$ and $v=r_{*}+t$, where
$r_{*}$ is the Regge-Wheeler ``tortoise'' co-ordinate defined by
$d/\,dr_{*}=(\Delta/r^{2})d/\,dr$. The co-ordinate $t$ is
spacelike between the event and the Cauchy horizons, and we take
$t=+\infty$ at the event horizon. In this letter we
are interested in the section $u=-\infty$ of the event horizon and the
section $v=+\infty$ of the inner horizon.
(These are the sections which intersect in the
standard Penrose diagram at future timelike infinity of the external
Universe.) We assume that the object moves along a typical radial
world-line that intersects the event horizon and the Cauchy horizon at
some finite values $v=v_{0}$ and $u=u_{0}$, respectively. Accordingly,
the trajectory of the object can be described by the function
$r(\tau )$ and by $u_{0}$, where $\tau$ is the proper time of the
infalling object. We set $\tau(r=r_{-})=0$. The details of $r(\tau)$ are
unimportant to our discussion. (The only piece of information that
enters the calculations is the value of $\dot{r}$ near $r_{-}$,
where a dot denotes differentiation with respect to
proper time. However, our results are not sensitive to this parameter.)

The class of perturbations that we consider here is the one which is
inherent to any non-spherical gravitational-collapse;
these are the electromagnetic
perturbations which result from the evolution of non-vanishing
electromagnetic
multipole moments (in the star) during the collapse. When these
perturbations propagate outwards, some fraction of them is
backscattered off the spacetime curvature and captured by the black hole.
This process leads to a ``tail'' of infalling radiation at the event
horizon which, at late times $(v\gg M)$, decays like $(v/M)^{-(2l+2)}$,
where $l$ is the multipole order of the mode \cite{price1}. This
electromagnetic field can be treated by the formalism developed in
\cite{gursel,chandrasekhar1,chandrasekhar2}.
We studied the asymptotic behavior of infalling electromagnetic
perturbations at the Cauchy horizon for polar modes \cite{burko2}.
Assuming $v_{0}\gg M$ (which also implies $-u_{0}\gg M$),
the divergent components of the Maxwell field strength tensor, as
expressed in the rest frame of an infalling observer, are given by the
following approximate expression to the leading order in
$\kappa_{-}\tau$ and $(\kappa_{-}u_{0})^{-1}$:
\begin{equation}
E=-B\approx C'(\kappa_{-}\tau)^{-1}\left( \ln|\kappa_{-}\tau|
+\frac{1}{2}\kappa_{-}u_{0}+\ln |\dot{r}|\right) ^{-(2l+3)},
\end{equation}
where $\kappa_{-}=(r_{+}-r_{-})/r_{-}^{2}$ is the surface gravity of
the Cauchy horizon;
$E(B)$ is the electric (magnetic) field, which points toward
the $\partial\,/\,\partial\theta$ $(\partial\,/\,\partial\phi )$
direction; $C'$ is a slowly varying function of $\theta$,$\phi$ (through
the Legendre polynomials) and is also proportional to the
initial value of the perturbing fields on the surface of the collapsing
star (or on the event horizon). The choice of the polar
modes is not expected to cause any loss of generality, since similar
qualitative behavior is to be anticipated for axial modes too.

The next stage of our analysis is to consider the interaction of the
divergent electromagnetic field $(1)$ with the matter comprising the
infalling object. We assume that the object is much smaller than the
radius of curvature between the event and the inner horizons, and hence
the effects of curvature are negligible. Consequently, we can
construe the object as being at rest in its locally co-moving Minkowski
frame when an electromagnetic impulse of the shape $(1)$ comes from null
infinity and interacts with it. Even in flat spacetime, the interaction
of matter with such an electromagnetic impulse is enormously complicated.
There are many
types of radiative processes, which may depend on the details of the
specific matter intricately. We therefore ignore all the
details of these radiative processes, and use a simplified toy-model to
describe the radiation--matter interaction.
Imagine that the object is made of
classical ``atoms''. (Later, we also consider a quantum analogue.)
Each ``atom" consists of two
electrically charged structureless particles with charges $+e$ and $-e$,
separated from each other by some internal force (e.g., a ``spring''). With
the lack of external forces, the system is static. In our case,
the  Lorentz force induced by the blue-shifted electromagnetic field
changes the separation between the two particles. Having in mind a small
deviation from equilibrium (which is justified {\it a posteriori}),
we assume a linear restoring force $F=-\mu\omega^{2}X$, where $X$ is
the deviation (of the particles' separation) from
equilibrium, $\omega$ is the resonance frequency, and $\mu$ is the reduced
mass. (The phrase ``atom'' refers here to an elementary unit of matter; we
do {\em not} consider a solar-type system here.)
The dipole is chosen to be aligned in the
$\partial\,/\,\partial\theta$ direction (to allow for a maximum
interaction with the field). We take the initial conditions to be $X=0$
and $\dot{X}=0$. The system's energy absorption is described by its
``excitation'', i.e., by the gain in kinetic and potential energy.
Although this model is extremely simplified, it may provide some insight
into the interaction of classical radiation with matter.

The equation of motion is $\mu\ddot{X}+\mu\omega^{2}X=eE(\tau)$,
where $E(\tau)$ is the divergent component of the electric field $(1)$.
(The contribution of the magnetic field is neglected, as the ratio of the
electric and the magnetic terms in the expression for the Lorentz
force is proportional to the system's internal velocity $\dot{X}$, which
is taken to be small --  a presumption which is justified
{\it a posteriori}.) The solution of this equation is
\begin{eqnarray}
X(\tau )=
-\frac{1}{2i\omega}e^{-i\omega\tau}\int_{-T}^{\tau }\frac{e}{\mu}
E(\tau ')e^{i\omega\tau '}\,d\tau '+{\rm c.c},
\end{eqnarray}
where $T$ is the time of infall from the event horizon to the Cauchy
horizon. It can be shown that the total absorbed mechanical energy of
the system ${\cal E}_{\rm c}$ up to proper time $\tau$ is
\begin{eqnarray}
{\cal E}_{\rm c}(\tau)=
\frac{1}{2}\mu\left|\int_{-T}^{\tau}\frac{e}{\mu}E(\tau ')
e^{i\omega\tau '}\,d\tau '\right|^{2}.
\end{eqnarray}
We assume that $\omega$ is of the order of magnitude of typical
molecular or atomic frequencies (or higher). For typical astrophysical
black holes we have $\omega M\gg 1$. This implies
$T\gg \omega^{-1}$. Hence, to evaluate the integral in $(2)$ and $(3)$
we divide the infall period into three
qualitatively different regions, denoted as regions $a,b$, and $c$,
respectively. In region $a$, defined by $-T< \tau\ll -\omega^{-1}$, the
slowly varying electric field can be taken outside the integration, and
therefore the absolute value of the
integral follows the electric field adiabatically. [This property means
that the system has no records of its past, so the results are insensitive
to the behavior of $E(\tau )$ at $\tau\ll -\omega^{-1}$.]
In region $c$, defined by $-\omega^{-1}\ll \tau < 0$, the
exponent in the integrand
can be taken outside the integral, and the remaining integration is
easily solvable. In between (region $b$), our assumption
$\omega^{-1}\ll M\ll -u_{0}$ implies that the variation in
the logarithmic term in $(1)$ is negligible throughout the region.
Taking this logarithmic term to be constant, the integral is easily
solvable for region $b$ too. Matching the solutions for the different
regions, it can be shown
\cite{burko1} that the total contribution of all three regions to
${\cal E}_{\rm c}$, $X$ and $\dot {X}$ on the Cauchy horizon is, to the
leading order in $(\kappa_{-}u_{0})^{-1}$,
\begin{eqnarray}
{\cal E}_{\rm c}(\tau=0)\approx\frac{1}{2(2l+2)^{2}}
\frac{C'^{2}}{\kappa_{-}^{2}}\mu\left(\frac{e}{\mu}\right)^{2}
\left(\frac{1}{2}\kappa_{-}u_{0}\right)^{-2(2l+2)}, \nonumber
\end{eqnarray}
\begin{eqnarray}
\dot{X}(\tau=0)\approx\frac{C'}{(2l+2)\kappa_{-}}\left(\frac{e}{\mu}
\right)\left( \frac{1}{2}\kappa_{-}u_{0}\right)^{-(2l+2)}, \nonumber
\end{eqnarray}
\begin{eqnarray}
X(\tau=0)\approx
-\frac{1}{2}\frac{\pi}{\omega}\frac{C'}{\kappa_{-}}\left(\frac{e}{\mu}\right)
\left( \frac{1}{2}\kappa_{-}u_{0}\right)^{-(2l+3)}.\nonumber
\end{eqnarray}
Now, it is easy to show that for fixed $r(\tau )$ there is a one-to-one
correspondence between
$v_{0}$ and $u_{0}$. In fact, $\,du_{0}/\,dv_{0}=-1$.
Therefore, the infalling observer can
increase $|u_{0}|$ by simply waiting outside the black hole before
jumping in and thus increasing the value of $v_{0}$. For a sufficiently
large $|u_{0}|$ we find that ${\cal E}_{\rm c}$, $X$ and $\dot{X}$ are
finite and small. Therefore, the behavior of the charged classical system
obtained by the above
analysis is regular, and although the external force acting on the system
diverges on the Cauchy horizon,
the energy absorbed by it is finite and negligible
for a sufficiently large $v_{0}$.

A quantum analogue can be conceived as a non-degenerate two-level system
(the ground state $|\psi_{i}\rangle$ and the excited state $|\psi_{f}
\rangle$) obeying Schr\"{o}dinger's equation. We take the system to be
initially in its ground state $|\psi_{i}\rangle$. The excitation of the
system can be described by the amplitude $a_{f}$ of the excited state
$|\psi_{f}\rangle$ in the wave-function.
This amplitude can be given \cite{burko1} in the Coulomb gauge by first
order time-dependent perturbation theory to be
\begin{eqnarray}
a_{f}(\tau)=-\frac{1}{\hbar\Omega_{fi}}\frac{e}{m}
\left\langle \psi_{f}\right|\left(\int_{-T}^{\tau}e^{i\Omega_{fi}\tau '}
E\,d\tau '\right)p\left|\psi_{i}\right\rangle  ,\nonumber
\end{eqnarray}
where $p$ is the component of the momentum 3-vector in the direction of
$\partial\,/\,\partial\theta$, $\hbar\Omega_{fi}$ is the energy
gap between the states, and $m$ is the electron mass. The integral here
is the same as in eq. $(3)$. It can then be shown
that on the Cauchy horizon the
absorbed energy ${\cal E}_{\rm q}(\tau)$
is, to the leading order in $(\kappa_{-}u_{0})^{-1}$,
\begin{eqnarray}
{\cal E}_{\rm q}(0)=\frac{1}{(2l+2)^{2}}\frac{1}
{\hbar\Omega_{fi}}\left(\frac{e}{m}\right)^{2}\frac{{C'}
^{2}}{\kappa_{-}^{2}}\left|\langle\psi_{f}|e^{-i\Omega_{fi}\bar{z}}p |
\psi_{i}\rangle\right|^{2}\left(\frac{1}{2}\kappa_{-}u_{0}\right)
^{-2(2l+2)},\nonumber
\end{eqnarray}
where $\bar{z}$ is a spatial Cartesian co-ordinate in the system's rest-
frame, pointed in the radial direction. The treatment here can be
generalized to a many level system. There is a remarkable correspondence
between our results for the classical system and for the quantum system as
can be clearly seen from the similarity of the expressions for the
absorbed energies in the two cases.

We have shown (in both the classical and the quantum models) that in
spite of
the divergence of the radiation's energy density (or even its integral
over proper time) as a consequence of the infinite blue shift, an
infalling observer may experience
just a finite effect upon crossing the Cauchy horizon.
Moreover, it is possible
to reduce the extent of that effect to a negligible impact, depending
only on how long after the collapse generating the black hole the observer
jumps into it. (Note, however, that physical effects {\em beyond}
the Cauchy horizon are as yet unknown.)

Obviously, the toy-model used here for the radiation--matter interaction
is a very simplified one. We believe, however, that
this model captures the main essence of the problem. We also note that
our treatment here is based on a first order perturbation analysis; yet
we do not expect higher-order contributions to change the qualitative
picture significantly.

We have not considered here possible {\sc Qed}
effects. In addition, a more
realistic model should consider other sources of electromagnetic fields
(e.g., the cosmic background radiation). After the completion of this
research we became aware of the the possibility that these effects may
have a much stronger impact than the
classical electrodynamic effects studied here. We are especially worried
about pair-production due to the interaction of the infalling matter with
the highly blue-shifted cosmic-radiation photons. Preliminary estimates
suggest to us that this process could be fatal for a human-being observer
(due to his high vulnerability to $\gamma$-rays),
but typical physical objects of similar or smaller size
might survive it. These effects are currently under investigation.

Further details of this work will be published elsewhere \cite{burko1}.

This research was supported in part by The Israel Science Foundation
administrated by the Israel Academy of Sciences and Humanities, the Fund
for the Promotion of Research at the Technion, and the Technion V.P.R Fund.

%\references

\end{document}